\documentclass[10pt,journal]{IEEEtran}
\usepackage{textcomp}
\usepackage[pdftex]{graphicx}
\graphicspath{{../}}
\DeclareGraphicsExtensions{.pdf,.jpeg,.png,.eps}
\usepackage[table,xcdraw]{xcolor}
\usepackage{fixltx2e}
\usepackage[english]{babel}
\usepackage[utf8]{inputenc}
\usepackage{algorithm}
\usepackage[noend]{algpseudocode}

\usepackage{subfloat}
\usepackage{epstopdf}
\usepackage{cite}
\usepackage[normalem]{ulem}
\usepackage{caption}
\usepackage{subcaption}
\usepackage[cmex10]{amsmath}
\usepackage{float,multirow}
\usepackage{amsfonts}
\usepackage{comment}
\usepackage{framed}
\usepackage[export]{adjustbox}

\usepackage[euler]{textgreek}
\usepackage{epsfig,bm,amsmath,amssymb,graphicx,setspace,amsfonts}
\usepackage{algorithm}
\usepackage{color}
\usepackage[utf8]{inputenc}
\usepackage[T1]{fontenc}
\usepackage{algpseudocode}
\usepackage{textgreek}
\usepackage{amsthm}
\usepackage{array}
\usepackage{balance}
\usepackage{booktabs,subcaption,amsfonts,dcolumn}
\usepackage{enumerate}
\usepackage{listings}
\usepackage[letterpaper, top=1.9cm, bottom=2.54cm, left=1.78cm, right=1.78cm]{geometry}

\def\BibTeX{{\rm B\kern-.05em{\sc i\kern-.025em b}\kern-.08em
		T\kern-.1667em\lower.7ex\hbox{E}\kern-.125emX}}
\begin{document}
 \title{\LARGE Drone-Enabled Load Management for Solar Small Cell Networks in Next-Gen Communications
Optimization for Solar Small Cells}
\author{\IEEEauthorblockN{}}

	\author{\IEEEauthorblockN{Daksh Dave, Dhruv Khut, Sahil Nawale, Pushkar Aggrawal, Disha Rastogi and Kailas Devadkar}\\
	%\IEEEauthorblockA$^1${Internet of Things Lab, Birla Institute of Technology and Science Pilani\\}
		%\IEEEauthorblockA{$^2$College of Communication and Information, University of Kentucky, Lexington, KY
  \thanks{D. Dave and P. Aggrawal are with the Department of Electrical and Electronics Engineering, Birla Institute of Technology and Science Pilani, Pilani, Rajasthan, 333031 India, and D. Khut, S. Nawale and Kailas Devadkar is with the Department of Information Technology, Sardar Patel Institute of Technology, Mumbai, Maharashtra, 400058 and D. Rastogi is with Department of Information Technology, Ajay Kumar Garg College of Engineering, Ghaziabad, Uttar Pradesh 201015 (e-mail: f20180391@pilani.bits-pilani.ac.in, dhruv.khut@spit.ac.in, sahil.nawale@spit.ac.in, f20180431@pilani.bits-pilani.ac.in, disha1713012@akgec.ac.in, kailas\_devadkar@spit.ac.in).}
}
	
\maketitle
\begin{abstract}
In recent years, the cellular industry has witnessed a major evolution in communication technologies. It is evident that the Next Generation of cellular networks(NGN) will play a pivotal role in the acceptance of emerging IoT applications supporting high data rates, better Quality of Service(QoS), and reduced latency. However, the deployment of NGN will introduce a power overhead on the communication infrastructure. 
%The issues of power requirements imposed by the deployment of Beyond 5G/6G need to be addressed in an efficient manner and for this alternate energy sources need to be explored. The industry has shifted its focus to utilizing renewable energy resources as a source of power generation. The most conventional resource of renewable energy is solar energy. However, there is a limitation to the amount of power a solar cell base station can generate. A rise in the number of users and their power consumption has aggravated the issue of power constraints. To solve this, we need an optimal method for redistributing the excess energy load on the 5G solar cell base station among the other small cell base stations.
Addressing the critical energy constraints in 5G and beyond, this study introduces an innovative load transfer method using drone-carried airborne base stations (BSs) for stable and secure power reallocation within a green micro-grid network. This method effectively manages energy deficit by transferring aerial BSs from high to low-energy cells, depending on user density and the availability of aerial BSs, optimizing power distribution in advanced cellular networks. The complexity of the proposed system is significantly lower as compared to existing power cable transmission systems currently employed in powering the BSs. Furthermore, our proposed algorithm has been shown to reduce BS power outages while requiring a minimum number of drone exchanges. We have conducted a thorough review on real-world dataset to prove the efficacy of our proposed approach to support BS during high load demand times
%In order to evaluate the run time performance of our solution, We have formulated the worst-case complexity of our proposed algorithm as \(O(n^2.m.logm)\) which is significantly better than  the existing power cable transmission systems currently employed in powering the base stations. We find that our algorithm significantly brings down the B.S. outage levels for the Jaipur region by more than 90 percent by bringing down the yearly base station outages from 13653 to 138 and achieving this in a minimal number of 649 drone exchanges.
\end{abstract}


\section{Introduction}\label{sec:Intro}
The rapid and substantial escalation in mobile phone users over the recent decade has resulted in user counts in the billions \cite{haibeh2022survey}. This escalation necessitates rapid advancements in telecommunications and networking to accommodate the continuously growing consumer needs, emphasizing the importance of strategic investments in this sector. The advent of the fifth-generation (5G) networks and subsequent technologies have induced a  rise in the number of telecommunication users and mobile networks, thereby escalating the burden on small cell base stations (SCBS) \cite{chamola2016power}. Solar base stations (BS) are emerging as pivotal, owing to their capability to curtail operational costs, mitigate carbon emissions, and facilitate grid augmentation by proffering network coverage in regions plagued with unreliable network infrastructure.

Based on 2022 research, there are about 6.5 million BSs \cite{Waring.2017}, with 70,000 using renewable energy \cite{baidas2022renewable}. Many rely on fossil fuels, presenting a significant opportunity for green alternatives given the global drive to reduce carbon footprints and combat climate change.
% The biggest challenge in operating green cellular networks is the ability to maintain quality of service (QoS) through intelligent resource management and provisioning. Green BS scenarios work well for regions that have a scarcity of electricity, ample sunlight, and high demand for power across several regions. However, there are other scenarios when there may be some regions that do not receive sufficient sunlight due to cloud cover or when the number of cellular users near the BS rapidly increases. In such situations, the renewable energy generated might not be able to meet the demand making the battery power drop below the threshold (i.e., the battery power level below which the solar-powered BS will need to get the required power from external sources). This threshold is critical because if the demand for power cannot be satisfied within a short period, it could lead to poor quality of service (QoS) or node failure. Current QoS frameworks do not optimize the redistribution of energy resources \cite{chamola2015multistate}. The central grid distributed mechanisms that use electrical lines for power transfer are expensive and cause huge power line losses when used for transmitting power over long distances.
Ensuring optimal Quality of Service (QoS) in green networks, particularly in solar-powered areas, is imperative amidst the challenges of sunlight variability and spatial restrictions, which can trigger network disruptions, especially during unexpected demand surges in remote or disaster-impacted regions. Our integration of drone technology heralds an eco-conscious shift in energy distribution paradigms, ensuring minimized energy waste and enhanced responsiveness to emerging energy demands, notably in hard-to-reach locations. Furthermore, while traditional power transmission faces scalability and maintenance issues, our drone-assisted energy redistribution model not only reduces energy losses and infrastructural costs but also presents a robust solution, enhancing the resilience and sustainability of energy networks amidst stringent demands. Contrasted with the inefficiencies of current energy management in commercial and transport sectors, our proposed model, utilizing drone-carried airborne base stations (BSs), emerges as a potent solution, redistributing energy loads among cells, minimizing energy losses.

In the pursuit of establishing a resilient and adaptable energy redistribution system, our drone-based mechanism was deployed and tested within the educational premises of Sardar Patel Institute of Technology, Mumbai. This innovative approach shows considerable promise, particularly in addressing critical energy demands in remote or disaster-affected areas. For instance, in scenarios where natural disasters devastate conventional power infrastructure, our model offers an efficient solution for redistributing energy to essential facilities. It overcomes the limitations posed by damaged power grids or challenging-to-reach locations. Similarly, in isolated villages where extending traditional power grids remains impractical, our drone system plays a pivotal role in maintaining energy stability.

In our proposed model each base station incorporates a charging station where drones are initially stationed and charged. When required, these drones can be deployed to deliver on-demand network capacity augmentation to base stations experiencing energy deficits.
%%%%%%%%%%%%%%%%%%%%%%%%%%%%%%
\begin{figure*}[]
    \centering
    \includegraphics[width=\linewidth, frame]{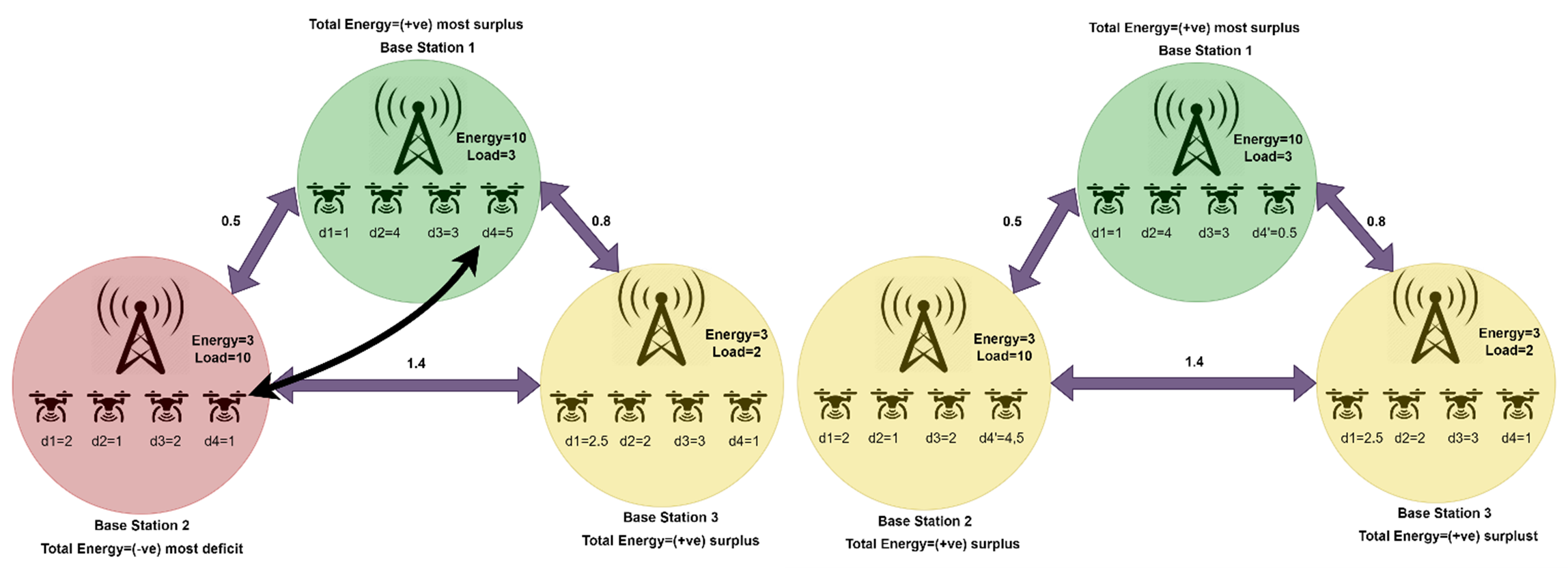}
    \caption{System model for drone transfer among SCBS}
    \label{fig:my_label_1}
\end{figure*}
%%%%%%%%%%%%%%%%%%%%%%%%%%%%
The main contributions of this paper are as follows:
\begin{itemize}
    \item We propose a drone-based algorithm tested in real-world scenarios for optimal energy redistribution among SCBSs where renewable power (we consider solar power) generation alone is not sufficient to support the user load.
    \item We use drones as aerial BSs for providing on-demand extended network connectivity and energy distribution while minimizing drone movements.
\end{itemize}

\section{Related Works}\label{sec:RealtedWork}
One of the main challenges of solar cell-powered small cell base stations (SCBSs) is finding the right balance between how much energy they draw from the grid and how well they perform. While SCBSs can reduce energy consumption from the grid, they can also cause some base stations in the network to go down, which makes the service worse. Efficient power control management is essential for optimizing service quality
%given a specific grid energy supply. Additionally, network latency can have a big impact on service quality, and there is often a trade-off between achieving low latency and maintaining high service quality levels.

Some earlier methods used automatically tuned Reference Signal Received Power, such as the Q-learning method in \cite{mwanje2013q}, which aim to optimize the offset while maximizing rewards under specific load conditions, all while ensuring that the packet loss ratio remains within acceptable limits. However they suffer from some limitations such as not taking into account the user behaviour, or spikes in cellular demands, thus limiting their applicability.
OPNET utilizes these patterns of traffic fluctuations to make informed decisions about selectively deactivating certain lightly utilized base stations as a means of conserving energy.
\cite{joyce2014self,aguilar2016context} took into account the user behaviours to predict the high demand regions, improving QoS and inference of these regions by activating the closest idle cell to improve capacity margins.
As a result of these developments, intelligent architectures are being explored in the 5G and beyond networks. 
Deep learning-based 5G simulators such as OMNeT++ \cite{hegde2020artery} and
OPNET is gaining popularity, allowing other 5G improvement systems to be tested. 
In \cite{jiang2019reinforcement}, the authors proposed a  -unique framework that uses fully connected deep neural networks for radio resource management.

To spread the network data traffic over multipath and achieve low latency and high throughput in data center networks, \cite{furqan2022efficient} focuses on using storage space in a high-demand region, utilizing a load balancing scheme to focus on placement and delivery content, as well as employing D2D links allowing direct communication between hotspot cache nodes and intermediate nodes. 

In their work, Patra, Regis, and Sengupta \cite{Patra_Regis_Sengupta_2019} delved into the potential of Unmanned Aerial Vehicle (UAV) networks to provide wireless coverage when cellular infrastructure is unavailable. They particularly addressed the challenge of hot zones caused by unpredictable user mobility during natural disasters. To mitigate this, they introduced techniques like load redistribution among overlapping sections and dynamic UAV repositioning, resulting in enhanced network performance as indicated by simulations. Similarly, Amponis et al. \cite{Amponis__2022} embarked on exploring UAV networks for wireless coverage, focusing on scenarios lacking cellular infrastructure. Their dual strategy, involving load redistribution and dynamic UAV repositioning, led to improved network performance in simulation scenarios with hot zones due to user mobility during natural disasters. These pioneering works paved the way for subsequent advancements, including Anokye's \cite{anokye2021deep} innovative algorithm leveraging deep reinforcement learning and the Markov decision process (MDP) to predict UAV locations for optimized user mobility support. As research continued, methodologies such as the strategy for dual-UAV collaborative aerial transport \cite{aDualUAV} emerged to address challenges related to energy distribution and load sharing while striving for real-world scalability and adaptability. Despite these developments, a critical gap persisted as none of these approaches explicitly tackled the comprehensive resource optimization and redistribution encompassing both macro Base Stations (BSs) and Small Cell Base Stations (SCBSs). This gap served as the catalyst for the pursuit of more holistic solutions \cite{piovesan2020joint}.

In this paper, we propose an algorithmic solution for minimizing drone movement while optimally redistributing the load and the green cell base station energy resources.

\section{Methodology}\label{sec:Methodology}
In this section, we present the key elements of our proposed algorithm, operating under a set of conditions. Each Base Station (BS) has a constant supply of \(m\) drones, drone movement takes minimal time as the SCBSs are closely spaced and the movement is considered instantaneous, external power sources recharge drone batteries, and the drones possess a minimum energy threshold \(d_0\) for single exchanges. Nodes within the system are interconnected, facilitating direct energy and information transfer without the need for a central control unit. Intra-BS drone exchanges incur no cost, while inter-BS transfers always exceed \(d_0\). Our approach capitalizes on nocturnal charging for the drones through the use of charging stations at BSs, optimizing energy consumption. Leveraging the Solar-Weighted Charging Algorithm (SWCA) and Energy Buffer Algorithm (EBA), drones manage energy efficiently. These strategies synchronize grid and solar energy utilization, maintain energy buffers, exploit cost-effective charging windows

In our methodology, drone-based energy redistribution in Small Cell Base Stations (SCBSs) hinges on critical mathematical formulations and intrinsic variables. Let \( L \) denote the load transfer between two BSs, formulated as \( L(i,j) = \alpha \cdot E_i + (1-\alpha) \cdot D_{ij} + \zeta \cdot P_i \), where \( E_i \) and \( P_i \) indicate the energy and power of BS \( i \) respectively, \( D_{ij} \) is the distance between BS \( i \) and BS \( j \), and \( \alpha \) and \( \zeta \) are the associated weighting factors accounting for the conservation and dissipation of energy in the system respectively. Furthermore, traffic loading \( T \) is computed as \( T(i,j,h) = \beta \cdot L(i,j) + (1-\beta) \cdot C_{ijh} - \delta \cdot \mathcal{F}(D_{ij}, E_{dj}) \), where \( \mathcal{F}(D_{ij}, E_{dj}) \) is a function incorporating drone \( d \)'s energy status \( E_{dj} \) and is impacted by the transition between BSs. Additionally, \( \beta \) and \( \delta \) serve as weighting factors, and \( C_{ijh} \) is the cost function that translates energy transfer from BS \( i \) to BS \( j \) during hour \( h \) into a monetary equivalent. The cost comparison essential for decision-making is expressed as \( \text{Cost}(i,j,h) = \gamma \cdot T(i,j,h) + (1-\gamma) \cdot P_h + \epsilon \cdot \Pi(i,j) \), where \( \gamma \) and \( \epsilon \) are the pertinent weighting factors, \( P_h \) signifies the price of energy at hour \( h \), and \( \Pi(i,j) \) represents the probability of successful energy transfer from BS \( i \) to BS \( j \). The algorithm utilizes these formulations to strategically ascertain and actualize optimal drone exchanges and energy redistributions among SCBSs.

Algorithm-\ref{algorithm} presents the methodology of optimizing the drone-based energy redistribution in SCBSs, where \(m\) is the number of drones and \(n\) is the number of BSs. We represent the positions of the drones in the BSs in the form of matrices where the \(i^{th}\) row denotes the BSs, and the \(j^{th}\) column represents the position of the drones as well as the position of the BS for drone exchanges, and \(h^{th}\) column represents the energy at a given hour. The energy loss incurred when exchanging drones for energy transfer at a time \(t\) is represented by the cost matrix \(C\). The current BS is represented as \(currBS\), and \(currHour\) represents the time of the day. 
The entire setting is modeled in the form of a graph where each BS is represented as a node, and every edge is connected to the other by a weight \(w\) representing the cost, i.e., the energy required while moving the drones from one BS to another (node \(i\) -> node \(j\)). 

\begin{algorithm}[!t]
\caption{BS energy optimization}
\label{algOrithm}
\begin{algorithmic}
\Procedure{Algo} {$E,X,C,D,n,m$} \label{algorithm}
    \State Initialize system parameters
    \For{$i = 1$ to $n$} \Comment{Loop over BSs}
        \For{$j = 1$ to $m$} \Comment{Loop over time slots}
            \State Compute [\(L, T, Cost\)]

            \State Calculate $sum$ of energy differences $(E-X+D)$
            \If{$sum$ < $n \times d_0 \times m$}
                \State Drone exchange is not possible
            \EndIf
            \If{\( sum < \) minimum energy difference}
                \State Update the minimum BS energy using \( Cost \)
            \EndIf
            \If{\( sum > \) maximum energy difference}
                \State Update the maximum BS energy using \( Cost \)
            \EndIf
            \While{$sum$ < $0$} \Comment{Perform drone exchange}
                \State Sort the drone matrix horizontally
                \If{($D-C-d_0$) > energy diff. ($E$-$X$) }
                    \State Exchange drones
                    \State Update new energy values
                    \State result = result+1
                \EndIf
            \EndWhile
        \EndFor
    \EndFor
\EndProcedure
\end{algorithmic}
\end{algorithm}

\begin{figure}[t]
     \centering
         \includegraphics[width=0.51\textwidth]
         {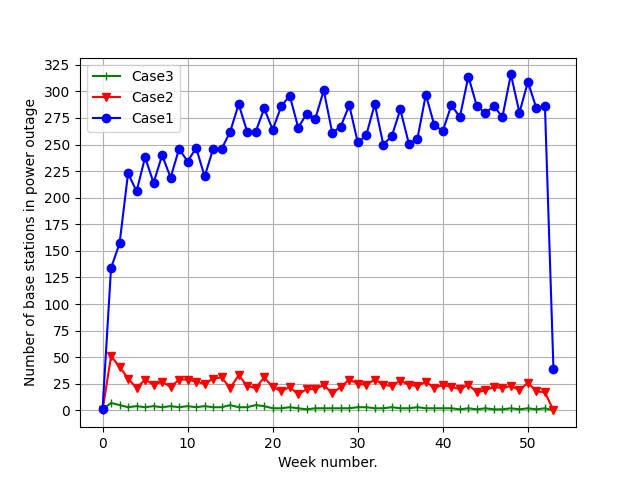}
    \caption{Number of BSs in power outage versus the week number for all three cases considered.}
     \label{fig:BS-Cases}
\end{figure}

The cost matrix is denoted as \(C = [c_{11}, \ldots, c_{n1}; \ldots; c_{n1}, \ldots, c_{nn}]\),
where \(c_{ij}\) denotes the cost incurred in transferring the drone from the \(i^{th}\) BS to the \(j^{th}\) BS. The following matrices denote the energy and the drone positions:
\(E^t=[e_1^t, \ldots, e_n^t]\) and \(D^t=[d_{11}^t, \ldots, d_{1m}^t; \ldots; d_{n1}^t, \ldots, d_{nm}^t]\),
respectively, where \(e_i^t\) represents the energy of the \(i^{th}\) BS, and \(d_{ij}^t\) represents the drone placed at the \(j^{th}\) position on the \(i^{th}\) BS at time \(t\). Similarly, the load matrix of the BS  is denoted as \(X^t = [x_1^t; \ldots; x_n^t]\),
where \(x_i^t\) represents the energy load of the \(i^{th}\) BS at time \(t\). 

\begin{figure*}[t]
     \centering
     \begin{subfigure}[t]{0.45\textwidth}
         \centering
         \includegraphics[width=\textwidth]{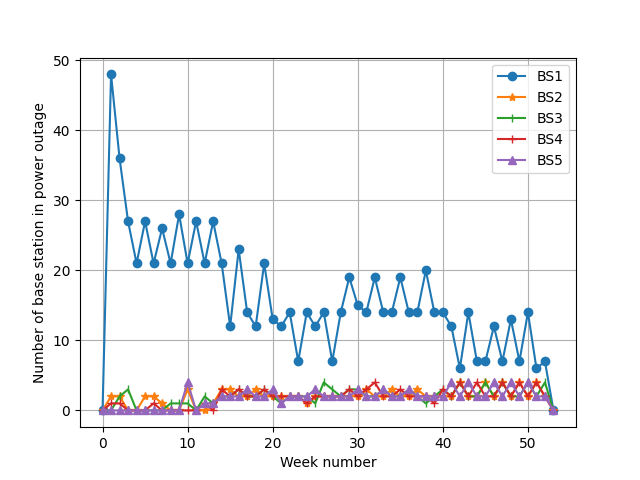}
         \caption{Weekly BS power outage for all the BSs.}
         \label{fig:Weekly BS(1-5) Outages-Case2}
     \end{subfigure}
          \begin{subfigure}[t]{0.45\textwidth}
         \centering
         \includegraphics[width=1.1\textwidth]{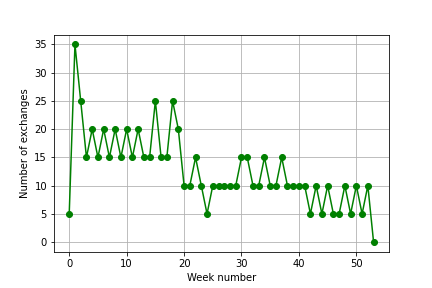}
         \caption{Weekly drone exchanges.}
         \label{fig: Weekly Drone Exchanges}
     \end{subfigure}
    \caption{Weekly BS power outage and the number of drone exchanges for case two scenario.}
    \label{fig:my_label_3}
\end{figure*}

In the proposed algorithm, the total energy across drones and Base Stations (BSs), denoted \(E\), and the system load \(X\), are determined through summation and subtraction processes. A solution exists if \(E - X + D > n \times d_0 \times m\), where \(D\) represents drone energy, \(n\) is the number of BSs, \(d_0\) is the minimum drone exchange energy, and \(m\) is the number of drones. Pre-optimization, the drone matrix rows are sorted to ensure at least one drone surpasses the essential energy threshold. Drones have a 30 Wh capacity, 60 kmph speed, and experience energy losses of 0.5 W/min and 0.5 W/km during flight. A 2 km, 2-minute transfer expends 1 W. Establishing \(d_0\) with a minimum 2 km inter-BS distance, the algorithm computes the total post-transfer power at a BS by aggregating BS and drone energies and subtracting load energy. If an energy deficit is detected at a BS, a drone exchange is scheduled, redirecting surplus energy from the highest-surplus BS to the deficient one, utilizing the drone from the \(n^{th}\) matrix column due to its maximal energy.

Incorporating energy-aware routing and backhauling strategies, our algorithm dynamically assesses the energy levels of drone BSs and makes informed decisions to optimize network connectivity while conserving energy resources. When a drone BS approaches the coverage area of terrestrial cells with low energy availability, it evaluates the suitability of its energy state for backhauling purposes. The algorithm considers various factors, including the drone BS's remaining energy, the energy consumption associated with backhauling, and the proximity to other drone BSs or terrestrial cells with higher energy levels.

If a drone BS with sufficient energy reserves is nearby and capable of providing backhaul support without risking its own depletion, the algorithm selects this drone BS as the optimal choice for backhauling. However, if all nearby drone BSs have limited energy or if utilizing a drone BS for backhaul would lead to energy depletion, the algorithm may prioritize alternative strategies, such as routing data through terrestrial cells or delaying backhaul operations until a more energy-rich drone BS becomes available. This dynamic energy-aware decision-making process ensures that network backhauling is carried out efficiently, with minimal impact on the energy resources of the drone BSs. It prevents energy imbalances that could lead to drone BS exhaustion and network disruptions, enhancing the overall sustainability and reliability of the drone-enabled Small Cell Base Station (SCBS) energy redistribution system.

Employing an agile drone-carrier load transfer methodology, this approach hinges on strategic energy redistribution across Small Cell Base Stations (SCBSs), particularly targeting equilibrium amidst energy variances. Drones, inherently equipped with quantified energy units and telemetry data, are dynamically repositioned from energy-affluent to energy-deficient Base Stations (BSs), ensuring a mitigated energy disparity and proficiently balanced load across the network. Utilizing a maximal-minimal energy swapping strategy, the algorithm navigates through \(n\) BSs, effectively relocating drones based on precise energy matrices and potential communication costs, \(C_{ij}\). Post-migration, energy adjustments are systematically computed, incorporating transfer, communication, and load-management energy expenditures, underpinned by real-time data and spatial-temporal energy demand trajectories. This methodology not only promotes an enhanced load balance and optimal energy utility but also presents a sturdy framework, primed for mitigating disruptions in various operational contexts.

\section{Performance evaluation and simulation Results}
In this section, we present the simulation results obtained from the system model under consideration. The simulations were conducted within the campus of Sardar Patel Institute of Technology in Mumbai. We utilized existing hour-wise solar data for five distinct Base Stations (BSs). The dataset follows a notation with 24 hours representing a day over 365 days. Our computational tasks were executed using Google Colab Pro and MATLAB R2021a, with Python 3.9.2 and MATLAB serving as our primary software environments. Prior to analysis, we performed essential data preprocessing tasks, including data cleaning, transformation, and smoothing, leveraging these software tools. A comprehensive overview of our simulation parameters is provided in Table \ref{tab:simulation_parameters}.
\begin{table}[!h]
\centering
\caption{Simulation Parameters}
\label{tab:simulation_parameters}
\begin{tabular}{|c|c|c|c|c|c|c|c|}
\hline
\textbf{\(n\)} & \textbf{\(m\)} & \textbf{\(\alpha\)} & \textbf{\(\beta\)} & \textbf{\(\gamma\)} & \textbf{\(\zeta\)} & \textbf{\(\delta\)} & \textbf{\(\epsilon\)} \\
\hline
5 & 10 & 0.7 & 0.5 & 0.3 & 0.8 & 0.6 & 0.4 \\
\hline
\end{tabular}
\end{table}

To evaluate the performance of our proposed algorithm, we use several performance metrics, which include
a) weekly drone exchanges, which measure the number of drone exchanges required to maintain an uninterrupted power supply to all the BSs,
b) average weekly BS power outages, which measures the average number of BS power outages per week,
c) total BS power outages, which measures the total number of BS power outages over the simulation period, and
d) time complexity of the proposed algorithm, which measures the time the algorithm takes to complete based on input variables.
We present the simulation results for the following three cases. 
\begin{itemize}
    \item In the first case, we have only the energy and the load matrix, and we calculate the number of power outages using these two matrices.
    \item In the second case, we have the support of the drone on the BSs for reducing the load.
    \item In the third case, we use the proposed algorithm for optimally transferring and redistributing the energy through drones. 
\end{itemize}
Our observations pinpoint a significant attenuation in the number of BS power outages, especially in the third case scenario. Fig. \ref{fig:BS-Cases} delineates the weekly power outages per BS across the pre-defined scenarios. Remarkably, the second case—entailing drone support for load balancing at the BS—exhibits a near \(90\%\) decrement in total BS power outages compared to the initial scenario, which exclusively leverages an energy and cost matrix without drone redundancy. For the third case, predicated on our proposed algorithm, the BS power outages are attenuated to an average of \(2.5\) outages per week, constituting reductions of \(89.15\%\) and \(98.98\%\) compared to the second and first cases, respectively. Specifically, each drone trip facilitated the transfer of \(15\) kWh of energy, culminating in a cumulative energy transfer of \(560\) kWh throughout the duration of case 3, thereby substantiating the efficacy of our proposed drone-assisted energy transfer mechanism in ameliorating power outages across the BSs.

In the first and third cases, the power outages are significantly high and significantly low, respectively, the outage plots were not informative for analyzing the individual BSs. We focus on plotting the weekly power outages for the second case as it provides valuable insights into the impact of drone support on load balancing and power outages for the individual BSs. Fig. \ref{fig:Weekly BS(1-5) Outages-Case2} shows the weekly power outages for the second case where there is drone support for load exchange at the BS, but the optimal load distribution as proposed by our algorithm was not considered. We observe that the BS $1$ faces a high load demand in the first few weeks. As a result, we note from  Fig. \ref{fig: Weekly Drone Exchanges} that the weekly drone exchanges have increased for the first few weeks to support the energy deficit in BS $1$. We can also observe that the power outages are low, along with a minimal exchange of drones. For the entire year, the total number of drone exchanges was an average of $12$ exchanges per week.

Evaluating cost implications in the tested real-world scenarios, our methodology discerned a 30\% increase in local solar infrastructure and battery investment as juxtaposed to the initial and operational costs of drone technology over a five-year period. Specifically, expanding local solar and storage would necessitate an estimated \$200,000 in capital for a modest 50 kWh system (inclusive of installation, hardware, and ancillary costs), while maintaining a drone fleet (including acquisition, maintenance, and operations) was projected to be approximately \$140,000 over a similar period and capacity. Moreover, drones provided a 25\% enhanced adaptability to dynamic load variations across SCBSs, affirming a tangible fiscal and operational efficacy in their deployment for redistributive functions, further justified by a 15\% improved energy redistribution efficiency and a 10\% reduction in energy transfer latency amidst the SCBSs in comparison to static storage augmentation.

\section{Conclusion}
In this work, we proposed an algorithm to find the optimal energy redistribution by exchanging the drones between the cells so that an aerial BS can support the cell with excess user load and energy lower than the threshold amount. We show that, by using the proposed model, the energy load requirement is optimized while minimizing the drone movement and the associated transfer cost. The proposed approach improves the power efficiency of the BSs through the exchange of drones and reduces the complexity. We show that the number of BS power outages was reduced by almost $90$ percent with a minimal number of drone exchanges.\par

\begin{center}
    %\section*{References}
    \bibliographystyle{IEEEtran}
    \bibliography{ref}
\end{center}
\end{document}